\documentclass[11pt,a4paper]{article}
\usepackage{amssymb, amsmath, amsfonts, amsthm, graphicx}
\usepackage{jcappub}
\usepackage{multirow}
\usepackage{hhline}
\usepackage{array}

\newcommand{\e}{{\rm e}}

\title{Kantowski-Sachs Einstein-Aether Scalar Field Cosmological Models: The Sequel}

\author[a]{S. Mohandas,}
\author[a]{R. J. van den Hoogen,}
\author[a]{D. Winters,}
\author[a]{M. Dala}

\affiliation[a]{Department of Mathematics and Statistics, \\
St. Francis Xavier University,\\
2323 Notre Dame Avenue, Antigonish, N.S., Canada}

\emailAdd{x2015jvq@stfx.ca}
\emailAdd{rvandenh@stfx.ca}
\emailAdd{x2017uza@stfx.ca}
\emailAdd{x2018upt@stfx.ca}

\abstract{
Utilizing the autonomous system of ordinary differential equations derived in \cite{VanDenHoogen:2018anx} to define the evolution, we further investigate a class of cosmological models within an Einstein-aether gravitational framework by introducing a non-trivial coupling between the shear of the aether field to the scalar field on the future asymptotic solution. We subsequently conduct qualitative and numerical stability analysis on the new set of equilibrium points and paramountly determine that the expansionary power-law inflationary attractor becomes anisotropic rather than isotropic in the presence of such a coupling. It is further shown that the stability of this solution is dependent on the value of the shear coupling parameter $a_3$. We also discover a family of asymptotically stable periodic orbits which exist for a particular range of parameter values within the Bianchi I invariant set and vanish in the absence of coupling between the aether field and the scalar field.}

\begin{document}
\maketitle
\section{Introduction}
\subsection{Lorentz-Violating Scalar Field Cosmological Models}

A notable obstacle for the advancement of General Relativity (GR) lies in its inability to harmonize with quantum mechanics and produce a consistent framework for a quantum theory of gravity. The potential prediction by notable quantum gravity theories of a preferred rest frame at the microscopic level in vacuum necessitates a violation of Lorentz invariance \cite{Liberati:2013xla,Jacobson:2008aj,Alfaro:2001rb,Gambini:1998it}. Further if one allows violations of Lorentz invariance, divergences in perturbative expansions of quantum field theories can be addressed \cite{Visser:2009fg}.  Hence, there are some theoretical reasons for investigating physical theories with Lorentz non-invariance to remedy these discordances. Here we are interested in investigating a Lorentz non-invariant gravitational theory containing a scalar field.

Conventional Einstein-aether theory consists of GR coupled to a dynamical, timelike unit vector field ($u^a$): the aether \cite{Jacobson:2000xp,Jacobson:2008aj,Jacobson:2004ts,Jacobson:2010mxa,Jacobson:2010mxb,Garfinkle:2007bk,Garfinkle:2011iw}. This aether field breaks Lorentz invariance but maintains local rotational symmetry and therefore only the boost sector of the Lorentz symmetry is broken. Typically, this Lorentz violation only occurs in the gravity sector of Einstein-aether theory, however, a logical extension is to permit similar Lorentz violations within the matter sector.  If the matter sector is described by a scalar field, then we can achieve a Lorentz violation if the scalar field is assumed to be coupled to the aether field within the scalar field potential \cite{Donnelly:2010cr,Barrow:2012qy,Sandin:2012gq,Solomon:2013iza,Kanno:2006ty,Solomon:2015hja,Alhulaimi:2017,Alhulaimi:2017ocb,VanDenHoogen:2018anx}. Should the scalar field be identified with the inflaton, this extension and effective aether coupling could result in the modification of standard inflationary dynamics found in General Relativity.

In GR, if the matter is described by a scalar field $\phi$, with an exponential potential $V=V_0e^{-2k\phi}$, then all {\em ever-expanding} spatially homogeneous models have an asymptotic power-law inflationary isotropic solution when the scalar field potential is sufficiently flat, that is when $k^2<\frac{1}{2}$ \cite{Halliwell:1986ja,Burd:1988ss,Kitada:1991ih,Kitada:1992uh}.  Of course, for a subset of the positive spatial curvature models of Bianchi type IX, Kantowski-Sachs, and closed FRW models, there exist models that are not {\em ever-expanding} when $k^2<\frac{1}{2}$ and recollapse \cite{Burd:1988ss,Kitada:1991ih,Kitada:1992uh}.  If $k^2>\frac{1}{2}$, then the only models that can possibly isotropize are those of Bianchi types I, V, VII, or IX \cite{Ibanez:1995zs,Coley:1997nk}. Bianchi type VII\textsubscript{h}  cosmological models with $k^2>\frac{1}{2}$ can isotropize towards an ever-expanding isotropic model, however, these asymptotic states are not power-law inflationary solutions \cite{vandenHoogen:1996vc}. Additionally, all initially expanding Bianchi type IX models with $k^2>\frac{1}{2}$ do not isotropize towards an ever-expanding isotropic model \cite{vandenHoogen:1998cc}.

Recently within an Einstein-aether gravitational framework, a class of spatially homogeneous but anisotropic cosmological models containing a scalar field which is coupled to the expansion $\theta=\nabla_au^a$ of the aether field through a generalized exponential potential $V(\phi,\theta)$ has been investigated \cite{Alhulaimi:2017,Alhulaimi:2017ocb,VanDenHoogen:2018anx}.  It has been observed that the existence of the coupling between the scalar field and the expansion of the aether field allows for the possibility of a late-time isotropic inflationary attractor for values of the parameter $k^2>\frac{1}{2}$ \cite{VanDenHoogen:2018anx}.

In this paper, we extend the analysis found in \cite{Alhulaimi:2017,Alhulaimi:2017ocb,VanDenHoogen:2018anx} by investigating the effects of coupling the scalar field to the shear scalar $\sigma$ of the aether field via a generalized scalar field potential of the form $V(\phi,\theta,\sigma^2)$. The complete derivation of the governing set of equations for the explicit model under consideration can be found in \cite{VanDenHoogen:2018anx}, therefore, for brevity we only summarize all pertinent equations and assumptions and refer the reader to \cite{VanDenHoogen:2018anx} for any missing details. The primary purpose of this paper is to determine the effects of coupling the shear of the aether field to the scalar field on the future asymptotic solution.

\subsection{Einstein-Aether Gravity with a Scalar Field}

The Lagrangian for Einstein-aether gravity (see \cite{Jacobson:2010mxa,Jacobson:2000xp,Jacobson:2010mxb,Garfinkle:2011iw,Garfinkle:2007bk,Jacobson:2008aj,Jacobson:2004ts,Donnelly:2010cr,Solomon:2013iza,Coley:2015qqa,Alhulaimi:2017ocb} for details) is
\begin{equation}
{\mathcal L}^{\textsc{AE}}=\frac{1}{2}R - K^{ab}_{\phantom{ab} cd}\nabla_{a}u^c\nabla_{b}u^d  + \lambda(u^au_a+1), \label{Lagrangian_AE}
\end{equation}
where
\begin{equation}
K^{ab}_{\phantom{ab}{cd}}  = c_1 g^{ab} g_{cd} + c_2\delta_{c}^{a} \delta_{d}^{b}+c_3\delta_{d}^{a}\delta_{c}^{b}+ c_4 u^{a} u^{b} g_{cd},\label{K}
\end{equation}
and where $\lambda$ is the Lagrange multiplier.
The Lagrangian for the scalar field is
\begin{equation}
L^\textsc{{M}}= -\frac{1}{2} g^{a b } \nabla_{a}\phi \nabla_{b}\phi - V(\phi,\theta,\sigma^2),
\end{equation}
where the scalar field potential is assumed to be a function of the scalar field, $\phi$, the expansion scalar, $\theta=\nabla_a u^a$, and the shear scalar $\sigma^2=\frac{1}{2}\sigma_{ab}\sigma^{ab}$, of the aether vector field.

Varying the total action
\begin{equation}
S=\int d^{4}x\sqrt{-g}\left[\frac{1}{8\pi G} {\mathcal L}^{\textsc{AE}}+{\mathcal L}^{\textsc{M}} \right] .\label{action}
\end{equation}
with respect to the inverse metric $g^{ab}$, the aether vector $u^a$, the scalar field $\phi$, and the multiplier $\lambda$ respectively yields the Einstein-aether, Klein-Gordon and the normalization field equations
\begin{eqnarray}
G_{ab}   &=& T_{ab}^{\textsc{U}} + 8\pi G \,T_{ab}^{\textsc{M}},\label{ae_equation}\\
-2\lambda u_a &=&  \frac{\delta {\mathcal L}^{\textsc{U}}}{\delta u^a}+8\pi G\,\frac{\delta {\mathcal L}^{\textsc{M}}}{\delta u^a},\label{deltaL-deltau}\\
\nabla^a\nabla_a\phi-V_{\phi}&=&0,\label{GeneralKleinGordoneqaution}\\
u^a u_a &=& -1.\label{Normalizationequation}
\end{eqnarray}
The effective energy momentum tensor due to the aether vector field is
\begin{eqnarray}
T_{ab}^{\textsc{U}} &=& 2\nabla_{c}\Bigl(J_{(a}^{\phantom{a}c}u^{\phantom{c}}_{b)} - J^c_{\phantom{c} (a}u^{\phantom{c}}_{b)} -J_{(ab)}u^c \Bigr) \nonumber\\
&& +2c_1\Bigl((\nabla_{a} u^c)(\nabla_{b} u_c) - (\nabla^c u_a)(\nabla_c u_b)\Bigr) -2c_4\dot{u}_a\dot{u}_b \nonumber\\
&& -2\left( u^d\nabla_c J^{c}_{\phantom{c}d}+c_4\dot{u}_c\dot{u}^c   \right)u_au_b - g_{ab} \Bigl(K^{cd}_{\phantom{cd}ef} \nabla_{c} u^e \nabla_{d} u^f\Bigr),\label{T_ab_AE}
\end{eqnarray}
where
\begin{eqnarray}
J^{a}_{\phantom{a}b}  &=& -K^{ac}_{\phantom{ac}bd} \nabla_{c} u^d,\\
\dot{u}^a &=& u^b\nabla_b u^a.
\end{eqnarray}
Additionally, the effective energy momentum tensor due to the scalar field is
\begin{eqnarray}
T_{ab}^{\textsc{M}} &=&  \nabla_a \phi \nabla_b \phi - \left(\frac{1}{2} \nabla_c\phi \nabla^c \phi + V \right)g_{ab}
%% The part from dependence on theta
+ \theta V_\theta g_{ab}+\dot{V}_\theta (g_{ab}+u_au_b) \nonumber\\
%% The part from dependence on sigma^2
&& +\left(\theta V_{\sigma^2}+\dot{V}_{\sigma^2}\right)\sigma_{ab}+V_{\sigma^2}\dot{\sigma}_{ab}-2\sigma^2V_{\sigma^2}u_au_b,
\label{scalar_T}
\end{eqnarray}
where $V=V(\phi,\theta,\sigma^2)$ and an overdot indicates derivative along the vector field $u^a$ via $\dot{f}\equiv\partial_a(f) u^a$.  The terms $V_\theta$ and $V_{\sigma^2}$ are the partial derivatives of the scalar field potential with respect to $\theta$ and $\sigma^2$, respectively.

\subsection{The Kantowski-Sachs Cosmological Model with a Scalar Field}

The geometry under consideration is spherically symmetric, spatially homogeneous and anisotropic represented by the Kantowski-Sachs metric
\begin{equation}
ds^2 = - dt^2 + a(t)^2 dx^2 + b(t)^2 (d\vartheta^2 + \sin^2 \vartheta  d\varphi^2).
\end{equation}
The aether vector field in co-moving coordinates is $u^{a}=(1,0,0,0)$, and the corresponding expansion and shear scalars are
\begin{equation}
\theta = \frac{\dot a}{a}+2\frac{\dot b}{b}, \qquad \sigma^2=3\sigma_+^2 =\frac{1}{3}\left(\frac{\dot b}{b}-\frac{\dot a}{a}\right)^2,
\end{equation}
while the shear tensor has the form $\sigma^a_{\phantom{a}b}=\mbox{Diag}[0,-2\sigma_+,\sigma_+,\sigma_+]$.

The effective energy density $\rho^{\textsc{U}}$, isotropic pressure, $p^{\textsc{U}}$, energy flux $q_{a}^{\phantom{a}\textsc{U}}$, and anisotropic stress $\pi_{\phantom{a}b}^{a\phantom{b}\textsc{U}}$   due to the aether field are:
\begin{eqnarray}
\rho^{\textsc{U}} & = & -\frac{1}{3}c_\theta \theta^2 -6c_\sigma \sigma_+^2,\\
p^{\textsc{U}} & = & \frac{1}{3}c_\theta \theta^2 +\frac{2}{3}c_\theta \dot\theta-6c_\sigma \sigma_+^2,\\
q_{a}^{\phantom{a}\textsc{U}}&=&0,\vphantom{\frac{1}{2}}\\
\pi_{\phantom{a}b}^{a\phantom{b}\textsc{U}}& = &2c_\sigma (\dot\sigma^a_{\phantom{a}b}+\theta\sigma^a_{\phantom{a}b}),\vphantom{\frac{1}{2}}
\end{eqnarray}
where new parameters $c_{\theta}= (c_1 +3c_2 +c_3)$ and $c_\sigma=(c_1+c_3)$ have been defined.
Based on astrophysical observations, the parameters $c_\theta$ and $c_\sigma$ satisfy the bounds \cite{Alhulaimi:2017ocb}
\begin{eqnarray}
0 &\leq& c_\sigma \leq \frac{1}{2}, \nonumber \\
1-2c_\sigma &\leq& \frac{1-2c_\sigma}{1+c_\theta} \leq 1-\frac{3}{2}c_\sigma. \label{C_constraints}
\end{eqnarray}
We note that the differential equations for $\dot \theta$ and $\dot \sigma_+$ reduce to algebraic equations when $1+c_\theta=0$ or $1-2c_\sigma=0$ respectively. Therefore for mathematical reasons, we assume that $1+c_\theta>0$ and $1-2c_\sigma>0$.  If we define $c^2=\frac{1-2c_\sigma}{1+c_\theta}$, then  $0< c \leq 1$ where $c=1$ reduces to conventional GR.

The generalized exponential scalar field potential which couples the expansion and shear scalars of the aether field to the scalar field is assumed to have the form
\begin{equation}
V(\phi, \theta, \sigma_+) = a_1 \e^{-2k\phi} + a_2\theta \e^{-k\phi} + a_3\sigma_+ \e^{-k\phi},\label{potential}
\end{equation}
where $k\geq0$ and $a_1 \geq 0$ are assumed.  The analysis in \cite{VanDenHoogen:2018anx} assumed that there was no coupling between the shear scalar of the aether field and the scalar field, $a_3=0$.  The primary purpose of this paper is to investigate a nontrivial coupling $a_3\not=0$.  The effective energy density $\rho^{\textsc{M}}$, isotropic pressure $p^{\textsc{M}}$, energy flux $q_{a}^{\phantom{a}\textsc{M}}$, and anisotropic stress $\pi_{\phantom{a}b}^{a\phantom{b}\textsc{M}}$ due to the scalar field are
\begin{eqnarray}
\rho^{\textsc{M}} & = & \frac{1}{2}\dot\phi^2+ a_1 \e^{-2k\phi},\\
p^{\textsc{M}} & = & \frac{1}{2}\dot\phi^2 -  a_1 \e^{-2k\phi} - ka_2 \dot\phi \e^{-k\phi} - a_3 \sigma_+ \e^{-k\phi},\\
q_{a}^{\phantom{a}\textsc{M}} & = & 0\vphantom{\frac{1}{2}}, \\
\pi^\textsc{M}_+& = & \frac{a_3}{6}\big(\theta - k\dot\phi\big) \e^{-k\phi},
\end{eqnarray}
where the anisotropic stress is $\pi_{\phantom{a}b}^{a\phantom{b}\textsc{M}}=\mbox{Diag}[0,-2\pi^\textsc{M}_+,\pi^\textsc{M}_+,\pi^\textsc{M}_+]$.

% ----------------------------------------------------------------
% --     Building the KS Model     -------------------------------
% ----------------------------------------------------------------

\subsection{The Field Equations in Normalized Variables}
In \cite{VanDenHoogen:2018anx} the normalized variables
\begin{equation}
\Phi = \frac{\sqrt{a_1}\e^{-k\phi}}{D}, \qquad
\Psi = \frac{\psi}{\sqrt{2}D},\qquad
  y = \frac{\sqrt{3}\sigma_+}{D},\qquad
  Q = \frac{\theta}{\sqrt{3}D},\label{variable1}
\end{equation}
where
\begin{equation}
D=\sqrt{\frac{1}{1+c_\theta}\frac{1}{b^2}+\frac{1}{3}\theta^2},\label{variable2}
\end{equation}
and a new time $\tau$ such that
\begin{equation}
\frac{d\tau}{dt}={D},
\end{equation}
were employed.  For ease of later analysis, we define new coupling parameters
\begin{equation}
\tilde{a}_2=\frac{a_2}{\sqrt{a_1}}, \mathrm{\ and\ } \bar{a}_3=\frac{\tilde{a}_3}{\sqrt{3}c}=\frac{a_3}{\sqrt{3a_1}c}.
\end{equation}
In which case equations \eqref{ae_equation}-\eqref{Normalizationequation} for the Einstein-aether Kantowski-Sachs geometry with a scalar field having a generalized exponential potential \eqref{potential} results in a four dimensional system of autonomous differential equations and a first integral which depend on four parameters $(k,c,\tilde{a}_2,\bar{a}_3)$
\begin{eqnarray}
\frac{d\Phi}{d\tau} &=&-\Phi\left(\sqrt{2}k\Psi+\chi\right),\label{dimensionlessDS1}\\
\frac{d\Psi}{d\tau} &=&-\sqrt{3}Q\Psi+k\Phi\left(\sqrt{2}\Phi+\sqrt{\frac{3}{2}}\tilde{a}_2Q+\frac{1}{\sqrt{2}}\bar{a}_3 c y\right)-\Psi\chi,\\
\frac{dy}{d\tau} &=&-\sqrt{3}Qy-\frac{1}{\sqrt{3}c^2}(1-Q^2)+\frac{\sqrt{3}\bar{a}_3}{c}\Phi\left(\frac{1}{2}Q-\frac{1}{\sqrt{6}}k\Psi\right)-y\chi,\\
\frac{dQ}{d\tau} &=& \frac{1}{\sqrt{3}}(1-Q^2)(Qy-\tilde{q}),\label{dimensionlessDS4}\\
f(\Phi,\Psi,y,Q)&=&1-c^2y^2-\Psi^2-\Phi^2=0, \label{dimensionlessconstraint}
\end{eqnarray}
where
\begin{equation}
\chi\equiv\frac{\dot D}{D^2}=\frac{1}{\sqrt{3}}\Bigl(y(Q^2-1)-Q(\tilde{q}+1)\Bigr),
\end{equation}
and
\begin{equation}
\tilde{q}\equiv qQ^2= 2c^2y^2 +2 \Psi^2 -\Phi^2-\frac{3}{\sqrt{2}}k\tilde{a}_2\Phi\Psi-\frac{3}{2}\bar{a}_3 c y\Phi.
\end{equation}
The deceleration parameter $q$ has the usual definition
\begin{eqnarray}
q   &=&- \left(3\frac{\dot{\theta}}{\theta^2}+1\right). \label{def_q}
\end{eqnarray}

Since the normalized variable $\Phi\geq 0$ by definition and given the constraint equation \eqref{dimensionlessconstraint} the dynamics of the variables $(\Phi,\Psi,y)$ are restricted to the upper half of the unit sphere surface.  In addition, we have
\begin{equation}
Q^2=\frac{1}{1+\frac{3}{(1+c_\theta)}\frac{1}{b^2\theta^2}}\leq 1,
\end{equation}
therefore, the phase space for the dynamical system is bounded and topologically equivalent to  $S^{2+} \times [-1,1]$.

%% ----------------------------------------------------------------
%% --     Qualitative Analysis      -------------------------------
%% ----------------------------------------------------------------
\section{Qualitative Analysis of the Dynamical System}

\subsection{The Invariant Sets}

There are three invariant sets determined by the curvature.  Orbits in the $Q^2<1$ invariant set are, in general, anisotropic and have positive spatial curvature, and therefore represent proper Kantowski-Sachs type models.  Orbits that lie in the $Q=\pm1$ invariant set represent anisotropic spatially homogeneous models having zero spatial curvature, and therefore comprise Bianchi type I models. Under the transformation
\begin{equation}
(\tau,\Phi,\Psi,y,Q,{\tilde{a}_2},\bar{a}_3)\to (-\tau,\Phi,-\Psi,-y,-Q,-{\tilde{a}_2},-\bar{a}_3), \label{mapping}
\end{equation}
the dynamical system \eqref{dimensionlessDS1}-\eqref{dimensionlessDS4} is invariant. Therefore the dynamics in the $Q=-1$ invariant set can be easily determined from the dynamics in the $Q=+1$ set via a time reversal.

The $\Phi=0$ invariant set represents massless scalar field models.  The dynamics in this set are independent of the coupling between the aether field and the scalar field and have been completely described in \cite{VanDenHoogen:2018anx}.

\subsection{The Equilibrium Points}

The equilibria for the system \eqref{dimensionlessDS1}-\eqref{dimensionlessDS4} can be categorized according to the invariant set ($Q^2<1$, $Q=+1$, or $Q=-1$) in which they lie.  The superscript indicates whether the point is in the $Q=+1$ or $Q=-1$ invariant set, with no superscript indicating the point is in neither. All of the equilibrium points of the dynamical system \eqref{dimensionlessDS1}-\eqref{dimensionlessDS4} are summarized in Table \ref{Table1}.

\begin{table}[ht]\renewcommand{\arraystretch}{2.4}
\begin{center}
\begin{tabular}{|c|c|}
\hline
Label & Values for $(\Phi,\Psi,y,Q)$  \\
\hline
$KS1_\delta$ &  $\displaystyle\left(0,0,  \frac{\delta}{c}, 2\delta c\right)$    \\
$KS2_\delta^{\phantom{\delta}\dag}$ &  $\displaystyle\left(\Phi^*,\Psi^*,y^*,Q^*\right)$    \\
$C^{+}$ &  $\displaystyle\left(0,\cos(u), \frac{\sin(u)}{c}, 1\right)$     \\
$C^{-}$ &  $\displaystyle\left(0,-\cos(u), -\frac{\sin(u)}{c},-1\right)$     \\
%% Bianchi I points
%%
$BI1_\delta^+$ & $\displaystyle\left(\frac{-2\sqrt{3}k\tilde{a}_2-\delta\bar{a}_3\sqrt{-\widetilde{K}}}{k(4+\bar{a}_3^{\,2})},
                               \,\,\frac{\sqrt{6}}{2k},
                               \frac{-\sqrt{3}k\tilde{a}_2\bar{a}_3+2\delta\sqrt{-\widetilde{K}}}{kc(4+\bar{a}_3^{\,2})},
                               1\right)$  \\
$BI1_\delta^-$ & $\displaystyle\left(\frac{2\sqrt{3}k\tilde{a}_2+\delta\bar{a}_3\sqrt{-\widetilde{K}}}{k(4+\bar{a}_3^{\,2})},
                               -\frac{\sqrt{6}}{2k},
                               \frac{\sqrt{3}k\tilde{a}_2\bar{a}_3-2\delta\sqrt{-\widetilde{K}}}{kc(4+\bar{a}_3^{\,2})},
                               -1\right)$  \\
$BI2_\delta^+$ &  $\displaystyle\left(
\frac{\sqrt{2}}{k\tilde{a}_2}\left(\Psi^{**}_+-\frac{\sqrt{2}}{\sqrt{3}}k\right),\phantom{+}\Psi^{**}_+,\frac{\bar{a}_3}{\sqrt{2}ck\tilde{a}_2}\left(\Psi^{**}_+-\frac{\sqrt{2}}{\sqrt{3}}k\right),
\phantom{+}1\right)$ \\
$BI2_\delta^-$ &  $\displaystyle\left(
\frac{\sqrt{2}}{k\tilde{a}_2}\left(\Psi^{**}_- +\frac{\sqrt{2}}{\sqrt{3}}k\right),-\Psi^{**}_-,\frac{\bar{a}_3}{\sqrt{2}ck\tilde{a}_2}\left(\Psi^{**}_- +\frac{\sqrt{2}}{\sqrt{3}}k\right),
-1\right)$ \\
\hline
\end{tabular}
\end{center}
\caption{Summary of all equilibrium points where $\delta=\pm 1$.  The points $KS2_\delta$ are missing in the corresponding table in reference \cite{VanDenHoogen:2018anx}.  The non-isolated circles of equilibrium points, $C^+$ and $C^-$, are parameterized by $u\in(-\pi,\pi]$.  We define the parameters $\widetilde{K}=3k^2\tilde{a}_2^{\,2}+(4+\tilde{a}_3^{\,2})\left(\frac{3}{2}-k^2\right)$ and $\Psi^{**}_\pm=k\left(\sqrt{2}(\bar{a}_3^{\,2}+4)\pm\delta2\tilde{a}_2\sqrt{\widetilde{K}}\right)\left(\sqrt{3}(\bar{a}_3^{\,2}+4+2k^2\tilde{a}_2^{\,2})\right)^{-1}$. The points $BI2_\delta^\pm$ reduce to isotropic models denoted as $FR_\delta^\pm$ in \cite{VanDenHoogen:2018anx} when $\bar{a}_3=0$. ($\dag$ See Appendix \ref{appendixa} for explicit form of $KS2_\delta$ equilibrium points)}\label{Table1}
\end{table}

\subsubsection{Equilibrium Points: \texorpdfstring{$Q^2<1$}{Qless1} -- Kantowski-Sachs}

\paragraph{Kantowski-Sachs: Vacuum Equilibrium Points - $KS1_{\delta}$}

The equilibrium points $$KS1_\delta=(0,0,  \frac{\delta}{c}, 2\delta c) \quad \mbox{where\ } \delta = \pm1,$$ represent Kantowski-Sachs anisotropic vacuum solutions with a positive deceleration parameter.
These points are in the phase space only if $c<\frac{1}{2}$, and therefore their existence is directly attributable to the existence of the aether field since general relativity has $c=1$.  The three curvature of the spatial hypersurface is $1-4c^2$ and the deceleration parameter is ${q}=(2c^2)^{-1}$ or $\tilde{q}=2$.  We note the following limits:
$$
\lim_{c\to\frac{1}{2}^-}\left(KS1_{+}\right) = {}^*C^+ \mbox{\qquad and \qquad} \lim_{c\to\frac{1}{2}^-}\left(KS1_{-}\right) = {}^*C^-,
$$
where ${}^*C^+$ and ${}^*C^-$ are special points on the circle of equilibria $C^+$ and $C^-$ respectively.

Calculating
$$\nabla f\big{|}_{KS1_\delta}=[0,0,-2c\delta,0],$$
indicates that we can eliminate the variable $y$ locally near the equilibrium points via the substitution $y=\frac{\delta}{c}\sqrt{1-\Phi^2-\Psi^2}$.
The eigenvalues for the resulting three dimensional dynamical system at the points $KS1_{\delta}$ are
$$\frac{\sqrt{3}\delta(1+2c^2)}{3c}, \frac{\sqrt{3}\delta(1-4c^2)}{3c},\frac{\sqrt{3}\delta(1-4c^2)}{3c}.$$
Therefore, the point $KS1_{+}$ which represents an expanding model, is a local source when it is in the physical phase space; i.e., $c< \frac{1}{2}$. Correspondingly, the point $KS1_{-}$ which represents a collapsing model, is a local sink when it is in the physical phase space; i.e., $c < \frac{1}{2}$. \cite{VanDenHoogen:2018anx}

\paragraph{Kantowski-Sachs: Scalar Field Equilibrium Points - $KS2_{\delta}$}

These equilibrium points are the Einstein-Aether generalizations of the scalar field Kantowski-Sachs equilibrium points found by Burd and Barrow in \cite{Burd:1988ss} for General Relativity. These points are also sometimes described as curvature scaling solutions \cite{Coley:2003mj}. Unfortunately, the equilibrium points $KS2_\delta$ are missing from the analysis in reference \cite{VanDenHoogen:2018anx} and therefore we shall expound upon them further here.

Assuming there is no coupling between the scalar field and the aether field for the moment, ($\tilde{a}_2=\bar{a}_3=0$), the equilibrium points can be expressed as
\begin{equation}
KS2_\delta=\left(\frac{\sqrt{2c^2+1}\sqrt{8c^2k^2-2k^2+3}}{\sqrt{3}\Xi},\frac{\delta\sqrt{6}k(2c^2+1)}{3\Xi},\frac{\delta(2k^2-1)}{\Xi},\frac{\delta(4k^2c^2+1)}{\Xi}\right)
\end{equation}
where
$(\Xi)^2=8c^4k^2+4c^2k^4+3c^2+1$. The deceleration parameter evaluated at this point is
\begin{equation}
q=\frac{2k^2-1}{4k^2c^2+1} \qquad\mbox{or}\qquad  \tilde{q}=\frac{(2k^2-1)(4k^2c^2+1)}{(\Xi)^2}.
\end{equation}
These points exist within the physical phase space when $k^2\leq \frac{1}{2}$ in which case the deceleration parameter is also negative. We also note that when ($\tilde{a}_2=\bar{a}_3=0$) then
\begin{eqnarray*}
{\lim}_{k^2\to{\frac{1}{2}}^-}\left(KS2_+\right) &=& {\lim}_{k^2\to{\frac{1}{2}}^-}\left( {}^{**}C^+\right), \\
{\lim}_{k^2\to{\frac{1}{2}}^-}\left(KS2_-\right) &=& {\lim}_{k^2\to{\frac{1}{2}}^-} \left( {}^{**}C^-\right),
\end{eqnarray*}
where ${}^{**}C^+$ and ${}^{**}C^-$ are special points on the circle of equilibria $C^+$ and $C^-$ respectively.

Given that $$\nabla f\big{|}_{KS2_\delta}\not=[0,0,0,0]$$ we can eliminate the variable $\Phi$ locally near the equilibrium point. The corresponding eigenvalues of the Jacobian matrix, $J$, evaluated at the equilibrium point $KS2_\delta$, are roots to a rather complicated cubic polynomial whose explicit form is not particularly illuminating.  However, given that the dimension of the system is 3, if the determinant of the matrix $J$ is positive then there is at least one positive eigenvalue and the point is unstable.  Further if the determinant of the matrix $J$ is negative and the trace of the matrix $J$ is positive then again there is at least one positive eigenvalue and again the point is unstable.

As we are not so much interested in the actual form of the determinant of $J$ and the trace of $J$ but in the sign of each, the analysis becomes manageable. We define
\begin{eqnarray}
D&\equiv&\mathrm{Sign}(Det(J))=\mathrm{Sign}\left(-\delta\left(2k^2-1\right)\right),\\
T&\equiv&\mathrm{Sign}(Tr(J)) =\mathrm{Sign}\left(-\delta\left(8c^2k^2-2k^2+3 \right)\right).
\end{eqnarray}
For the point $KS2_{-}$,  we find $D<0$ and $T>0$ and therefore the point is unstable with a 1-dimensional stable manifold.  For the point $KS2_+$, we observe $D>0$ and $T<0$ and therefore the point is also unstable with a 2-dimensional stable manifold. We can conclude that when there is no coupling between aether field and the scalar field, the points $KS2_\delta$ are unstable whenever they exist.

\begin{figure}[ht]
\begin{center}
	$\begin{array}{ccc}
     c=\frac{ 1}{32},\bar{a}_3=-\sqrt{2}  &  c=\frac{ 1}{32},\bar{a}_3=0  &  c=\frac{ 1}{32},\bar{a}_3=\sqrt{2}\\
	\includegraphics[width=0.3\textwidth]{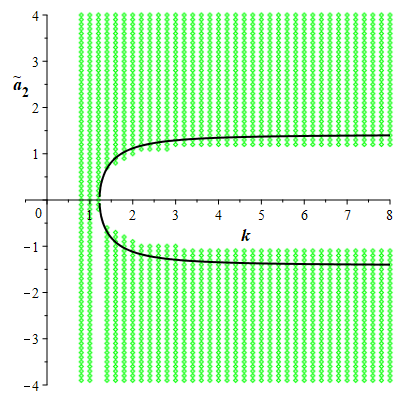}  &
    \includegraphics[width=0.3\textwidth]{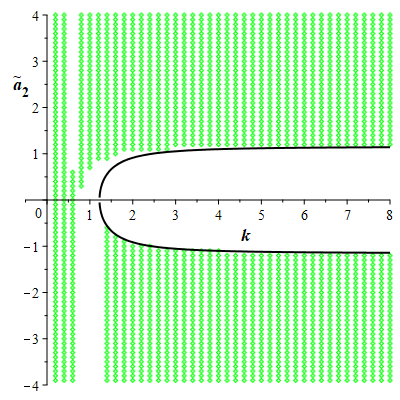}  &
	\includegraphics[width=0.3\textwidth]{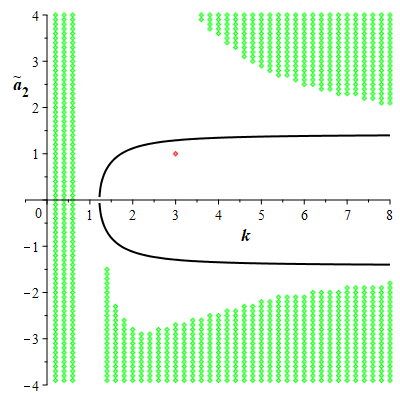} \\
    c=\frac{ 7}{32},\bar{a}_3=-\sqrt{2}  &  c=\frac{ 7}{32},\bar{a}_3=0  &  c=\frac{ 7}{32},\bar{a}_3=\sqrt{2}\\
	\includegraphics[width=0.3\textwidth]{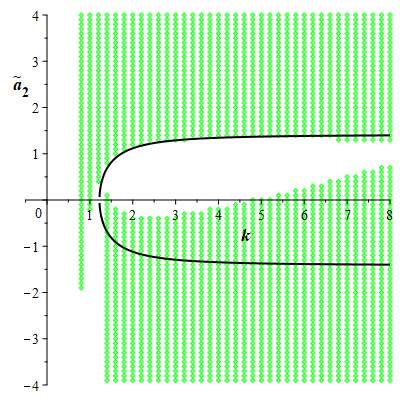}  &
    \includegraphics[width=0.3\textwidth]{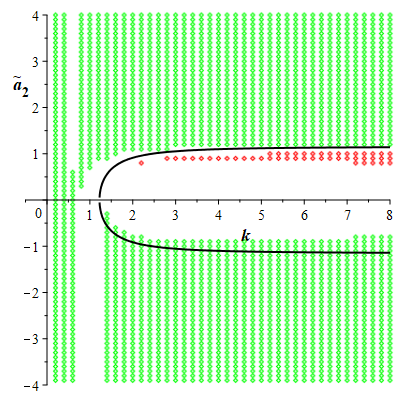}  &
	\includegraphics[width=0.3\textwidth]{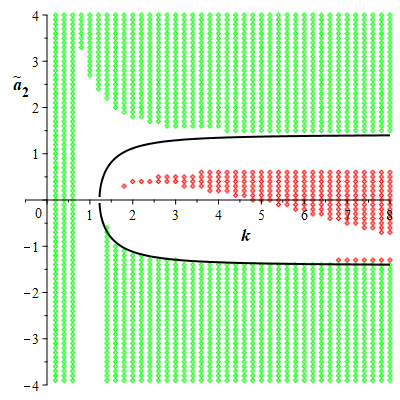} \\
    c=\frac{30}{32},\bar{a}_3=-\sqrt{2}  &  c=\frac{30}{32},\bar{a}_3=0  &  c=\frac{30}{32},\bar{a}_3=\sqrt{2}\\
    \includegraphics[width=0.3\textwidth]{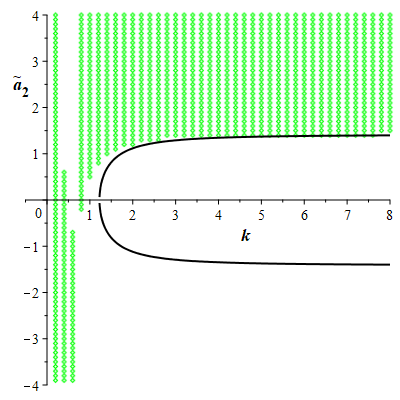}  &
    \includegraphics[width=0.3\textwidth]{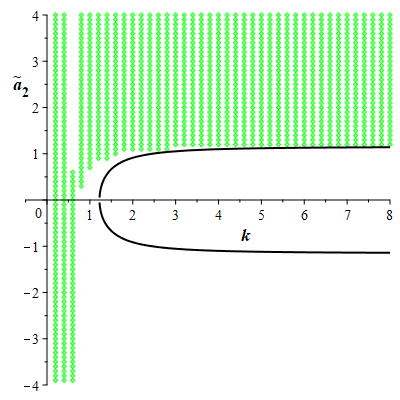}  &
	\includegraphics[width=0.3\textwidth]{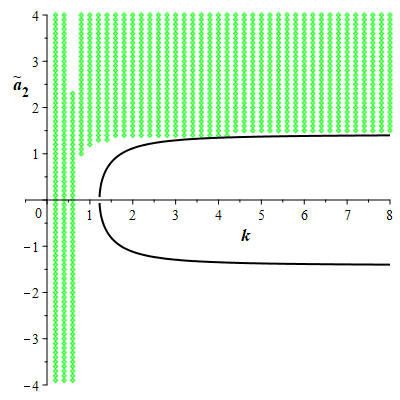}  \\
	\end{array}$
\end{center}
	\caption{The figures depicts stability for the $KS2_+$ equilibrium point.  Red indicates regions where the point is a sink, while blue indicates regions where the point is a source. Green indicates regions where the point is a saddle. The black curve represents $\tilde{K}=0$ and is included for reference only. }\label{KS2 stability9}
\end{figure}

\begin{figure}[ht]
\begin{center}
	$\begin{array}{ccc}
     c=\frac{ 1}{32},\bar{a}_3=-\sqrt{2}  &  c=\frac{ 1}{32},\bar{a}_3=0  &  c=\frac{ 1}{32},\bar{a}_3=\sqrt{2}\\
	\includegraphics[width=0.3\textwidth]{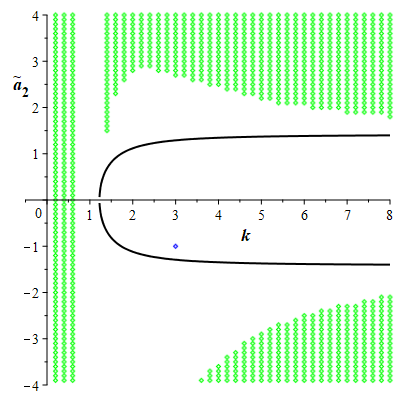}  &
    \includegraphics[width=0.3\textwidth]{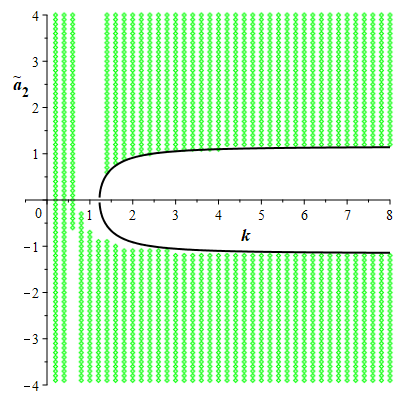}  &
	\includegraphics[width=0.3\textwidth]{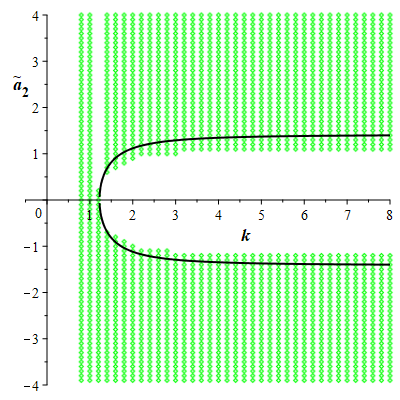} \\
    c=\frac{ 7}{32},\bar{a}_3=-\sqrt{2}  &  c=\frac{ 7}{32},\bar{a}_3=0  &  c=\frac{ 7}{32},\bar{a}_3=\sqrt{2}\\
	\includegraphics[width=0.3\textwidth]{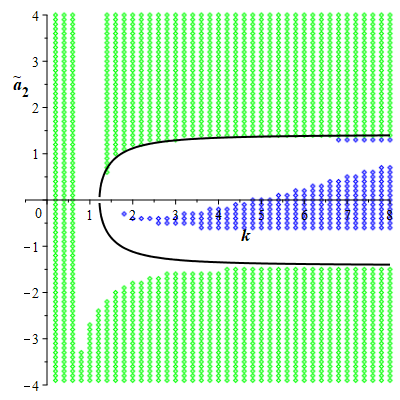}  &
    \includegraphics[width=0.3\textwidth]{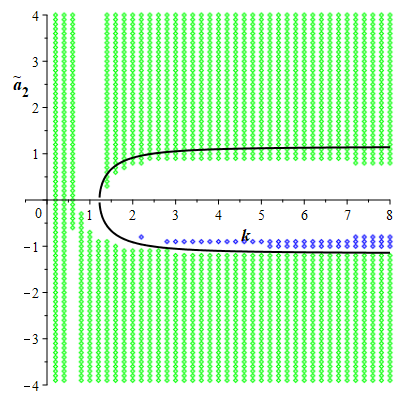}  &
	\includegraphics[width=0.3\textwidth]{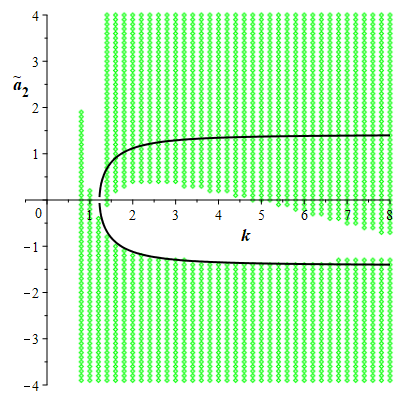} \\
    c=\frac{30}{32},\bar{a}_3=-\sqrt{2}  &  c=\frac{30}{32},\bar{a}_3=0  &  c=\frac{30}{32},\bar{a}_3=\sqrt{2}\\
    \includegraphics[width=0.3\textwidth]{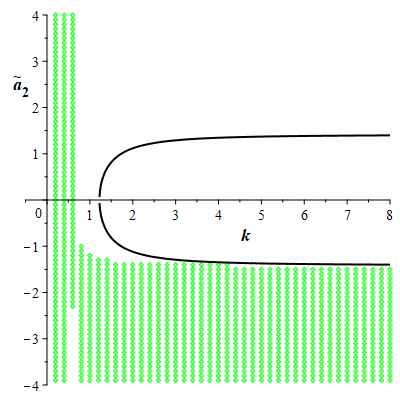}  &
    \includegraphics[width=0.3\textwidth]{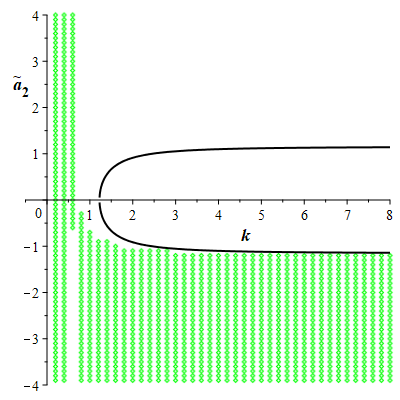}  &
	\includegraphics[width=0.3\textwidth]{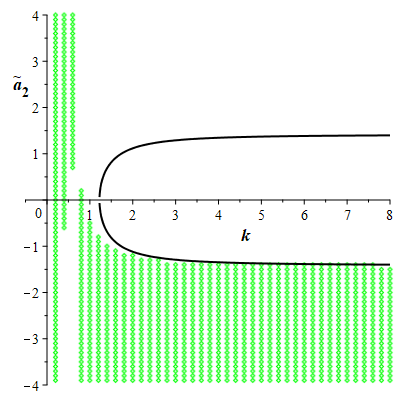}  \\
	\end{array}$
\end{center}
	\caption{The figures depicts stability for the $KS2_-$ equilibrium point.  Red indicates regions where the point is a sink, while blue indicates regions where the point is a source. Green indicates regions where the point is a saddle. The black curve represents $\tilde{K}=0$ and is included for reference only. }\label{KS2 stability10}
\end{figure}

If we include a nontrivial coupling between the scalar field and aether field, ($\tilde{a}_2\not=0$ or $\bar{a}_3\not=0$), then these points continue to exist in the phase space for various values of the parameters, including regions in which $k^2>\frac{1}{2}$. The explicit formulas for the equilibrium points in terms of the parameters are given in Appendix \ref{appendixa}.  Because of the two additional parameters, the calculation of the eigenvalues results in extremely complicated expressions which are not illuminating. Numerical experimentation can be employed to investigate the eigenvalues associated with these equilibrium points when $\tilde{a}_2\not=0$ and $\bar{a}_3\not=0$ (See Figures \ref{KS2 stability9} and \ref{KS2 stability10}). We observe that the equilibrium point $KS2_+$ can be either a saddle or a sink.  Whereas,  the equilibrium point $KS2_-$ can be either a saddle or a source. Other behaviours may be possible, since the full set of possible parameter values has not been fully explored.

\subsubsection{Equilibrium Points: \texorpdfstring{$Q=+1$}{Qis1}  -- Bianchi I}

\paragraph{Bianchi type I: Massless Scalar Field Equilibrium Points - $C^+$}

In the $Q=+1$ invariant set, there exists a circle of non-isolated equilibria given by
$$C^{+}=(0,\cos(u), c^{-1}\sin(u), 1) \mbox{\ with\ } u\in(-\pi,\pi],$$
which lies on the boundary of the phase space with a deceleration parameter $q=\tilde{q}=2$. Cosmologically, these points represent massless scalar field solutions and are analogues to the Kasner solutions. Since
$$\nabla f\big{|}_{C^{+}}=[0,-2\cos(u),-2\sin(u),0] $$
we can eliminate the variable $y$ locally at the equilibrium point. The corresponding eigenvalues for the resulting three dimensional system are
$$0,\sqrt{3}-\sqrt{2}k\cos(u),\frac{2\sqrt{3}(2c-|\sin(u)|)}{3c}.$$
The first two eigenvalues correspond to the dynamics within the $Q=1$ invariant set, with the zero eigenvalue indicating here the non-isolated nature of the set of equilibria.  In the full three-dimensional phase space the local stability of the non-isolated line of equilibria depends on parameters $k$ and $c$ and is independent of the coupling parameters $\tilde{a}_2$ and $\bar{a}_3$.  Full details of the stability for this line of equilibria is given in \cite{VanDenHoogen:2018anx}. If $c>\frac{1}{2}$ and $k^2<\frac{3}{2}$ then $C^+$ is a source. If $4c^2+\frac{3}{2k^2}<1$, then a part of $C^{+}$ is a source and there are two sections of $C^{+}$ that are sinks which are surrounded by saddles.   For the remainder of parameter space, a part of $C^+$ will always act as a source and a part will be a saddle.  In the two-dimensional $Q=1$ invariant set the asymptotic behaviour depends only on the value of $k^2$. Therefore, in the $Q=1$ invariant set, $C^+$ acts as a source if $k^2<\frac{3}{2}$ and if $k^2>\frac{3}{2}$ then part of $C^+$ is a source and part is a sink.

\paragraph{Bianchi type I: Coupling Dominated Equilibrium Points (type 1) - $BI1^+_{\delta}$}

The two points
\begin{equation}
BI1^+_{\delta} = (\Phi_{eq},\Psi_{eq},y_{eq},1)=\Bigg(\frac{-2\sqrt{3}k\tilde{a}_2-\delta\bar{a}_3\sqrt{-\widetilde{K}}}{k(4+\bar{a}_3^{\,2})},\frac{\sqrt{6}}{2k},\frac{-\sqrt{3}k\tilde{a}_2\bar{a}_3+2\delta\sqrt{-\widetilde{K}}}{kc(4+\bar{a}_3^{\,2})},1\Bigg)
\end{equation}
where $\delta=\pm1$, represent anisotropic expanding Bianchi type I models with a scalar field. The adjusted parameter $\tilde{K}$ is defined as
\begin{equation}
\tilde{K} = 3k^2{\tilde a}_2^{\,2}+\left(4+{\bar a}_3^{\,2}\right)\left(\frac{3}{2}-k^2\right)=K+{\bar a}_3^{\,2}\left(\frac{3}{2}-k^2\right)
\end{equation}
where $K=6+3k^2\tilde{a}_2^{\,2}-4k^2$ is the parameter used in \cite{VanDenHoogen:2018anx} when there was no shear coupling.  Since $q=\tilde{q}=2$ at these equilibrium points these points represent non-inflationary solutions. The solutions represented by these equilibrium points do not exist when there is no coupling between the scalar field and aether field since $\lim_{(\tilde{a}_2,\bar{a}_3)\to(0,0)} \left(BI1^+_\delta\right)$ approaches two isolated points on the circle of equilibria $C^+$.

\begin{figure}[ht]
\begin{center}
\includegraphics[width=0.4\textwidth]{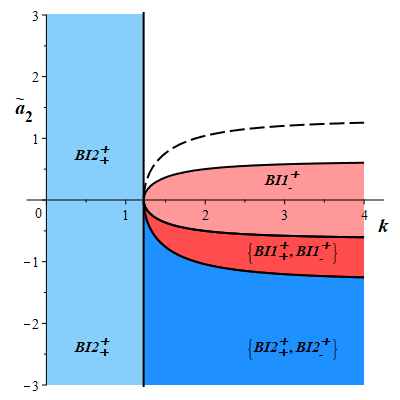}
\end{center}
\caption{The shaded areas of the figure indicate the portion of the parameter space $(k,\tilde{a}_2)$ in which the equilibrium points $BI1^+_\delta$ and $BI2^+_\delta$ exist in the physical phase space for $\bar{a}_3>0$. The dashed line and its symmetric continuance below the horizontal  axis marks the location of $\tilde K=0$ curve.  The vertical line is $k^2=3/2$. The darker regions indicate parameter values in which two points coexist, while blank regions indicate parameter values in which neither of these points exist. If $\bar{a}_3\to 0^+$ then the inner parabolic shaped curve becomes smaller and approaches the curve $\tilde{a}_2=0$ in which case there is no light red portion of the graph.  If $\bar{a}_3<0$, then the only change is in the light red region is $BI1^+_- \to BI1^+_+$.} \label{Existence_plot}
\end{figure}

The equilibrium points $BI1^+_\delta$ exist in the physical phase space when $\tilde{K}<0$ and $\Phi_{eq}\geq 0$ which corresponds to the parameter values
\begin{equation}
k^2\geq\frac{3}{2}, \mbox{\qquad  and\qquad } -\sqrt{\left(
\bar{a}_3^{\,2}+4\right)\left(\frac{1}{3}-\frac{1}{2k^2}\right)}<\tilde{a}_2<
-\delta\bar{a}_3\sqrt{\frac{1}{3}-\frac{1}{2k^2}}\ .
\end{equation}
See Figure \ref{Existence_plot} which summarizes graphically the existence of $BI1^+_\delta$.  We note the limiting case:
\begin{equation}
\lim_{\tilde{K}\to 0^-}\left(BI1^+_{\delta}\right) = \lim_{\tilde{K}\to 0^+}\left(BI2^+_{\delta}\right).
\end{equation}

Since
\begin{equation}
\nabla f\rvert_{BI1^+_\delta}=[-2\Phi_{eq},-2\Psi_{eq},-2c^2y_{eq},0],
\end{equation}
the variable $\Psi$ can be eliminated locally near each of the equilibrium points. The eigenvalues for the resulting three dimensional system in the variables $\Phi,y,Q$ at the points $BI1^+_\delta$ in terms of $y_{eq}$ are
\begin{eqnarray}
\lambda_{1,2}&=&\mathrm{Roots\ of\ }\left(\lambda^2-\frac{\tilde{K}}{4k^2(\bar{a}_3^{\,2}+4)} \left(4\sqrt{3}\tilde{a}_2\bar{a}_3ck^2y_{eq}
+6k^2\tilde{a}_2^{\,2}-\bar{a}_3^{\,2}(3-2k^2)\right)\right),\label{BIpluseigens}\\
\lambda_3&=&-\frac{2}{\sqrt{3}}(y_{eq}-2).
\end{eqnarray}
By simplifying the expression for the first two eigenvalues we can show that the result is a pair of pure imaginary eigenvalues since $\tilde{K}<0$;
\begin{equation}
\lambda_{1,2}=\pm i \sqrt{ \frac{-\tilde K}{2k^2(\bar{a}_3^{\,2}+4)^2} \left((12k^2\tilde{a}_2^{\,2}-\bar{a}_3^{\,2}\tilde{K})+
\mathrm{sgn}(\tilde{a}_2\bar{a}_3)\delta\sqrt{(12k^2\tilde{a}_2^{\,2}-\bar{a}_3^{\,2}\tilde{K})^2-\Gamma^2 }   \right)  }
\end{equation}
where $\mathrm{sgn}(x)=\frac{x}{|x|}$ and
\begin{equation}
\Gamma^2=\frac{1}{4}(\bar{a}_3^{\,2}+4)^2\left(\bar{a}_3^{\,2}(2k^2-3) -6 k^2\tilde{a}_2^{\,2}\right)^2.
\end{equation}
The eigenspace associated with the pair of imaginary eigenvalues lies in the $Q=1$ invariant set. The third eigenvalue is negative when $y_{eq}>2$. When this occurs, there exists periodic orbits in the neighborhood of $BI1^+_\delta$ that are asymptotically stable, and the point is called a stable center. The point $BI1^+_+$ is a stable center if
\begin{equation}
\tilde{a}_2\bar{a}_3<\frac{-2c(\bar{a}_3^{\,2}+4)}{\sqrt{3}} \mathrm{\quad or\quad } 4c^2+\frac{3}{2k^2}+\left(\frac{\sqrt{3}}{2}\tilde{a}_2+c\bar{a}_3\right)^2<1,
\end{equation}
while the point $BI1^+_-$ is an asymptotically stable center if
\begin{equation}
\tilde{a}_2\bar{a}_3<\frac{-2c(\bar{a}_3^{\,2}+4)}{\sqrt{3}} \mathrm{\quad and\quad } 4c^2+\frac{3}{2k^2}+\left(\frac{\sqrt{3}}{2}\tilde{a}_2+c\bar{a}_3\right)^2>1.
\end{equation}
Careful analysis of these two curves within the region of existence yields regions of parameter space in which the points $BI1^+_\delta$ are asymptotically stable centers.  These regions are described in Table \ref{Stability-table1}.

\begin{table}[ht]
$$\arraycolsep=1.4pt\def\arraystretch{2.2}
\begin{array}{|>{\centering\arraybackslash$} p{3.0cm} <{$}   >{\centering\arraybackslash$} p{3.0cm} <{$} >{\centering\arraybackslash$} p{4cm} <{$} >{\centering\arraybackslash$} p{3.0cm} <{$} |}
\hline
\bar{a}_3 & c^2 & k^2 & \tilde{a}_2\\
\hhline{|====|}
\multicolumn{4}{|l|}{{\bf BI1^+_+} }  \\
\bar{a}_3>0 & 0 < c^2 < \frac{\bar{a}_3^{\,2}}{4(\bar{a}_3^{\,2}+4)}           & \frac{3}{2(1-4c^2)}< k^2 < (k^*_1)^2   & D^*_{1-} < \tilde{a}_2 < \tilde{\tilde{K}}^*_- \\
\bar{a}_3>0 & 0 < c^2 < \frac{\bar{a}_3^{\,2}}{4(\bar{a}_3^{\,2}+4)}           & (k^*_1)^2 < k^2 < \infty                  & \tilde{K}^*_- < \tilde{a}_2 < \tilde{\tilde{K}}^*_- \\
\bar{a}_3>0 & \frac{\bar{a}_3^{\,2}}{4(\bar{a}_3^{\,2}+4)} < c^2 < \frac{1}{4} & \frac{3}{2(1-4c^2)}< k^2 < \infty       & D^*_{1-} < \tilde{a}_2 < \tilde{\tilde{K}}^*_- \\
\bar{a}_3=0 & 0 < c^2 < \frac{1}{4}                                            & \frac{3}{2(1-4c^2)} < k^2 <\infty       & D^*_{1-} < \tilde{a}_2 < 0 \\
\bar{a}_3<0 & 0 < c^2 < \frac{1}{4}                                            & \frac{3}{2(1-4c^2)}< k^2 < \infty       & D^*_{1-} < \tilde{a}_2 < \tilde{\tilde{K}}^*_+ \\
\hhline{|====|}
\multicolumn{4}{|l|}{{\bf BI1^+_-} }   \\
\bar{a}_3>0 & 0 < c^2 < \frac{\bar{a}_3^{\,2}}{4(\bar{a}_3^{\,2}+4)}        & (k^*_1)^2 < k^2 < \infty                  & \tilde{K}^*_-< \tilde{a}_2 < D^*_{1-} \\
\hline
\end{array}$$
\caption{Regions of the parameter space where the equilibrium point $BI1^+_\delta$ is an asymptotically stable center.  \newline
$\displaystyle (k^*_1)^2 =  \frac{3/2\bar{a}_3^{\,2}}{\bar{a}_3^{\,2}-4c^2(\bar{a}_3^{\,2}+4)}$, $\tilde{K}^*_{\pm} = \pm\sqrt{(\bar{a}_3^{\,2}+4)\left(\frac{1}{3}-\frac{1}{2k^2}\right)}$, $\tilde{\tilde{K}}^*_\pm = \pm|\bar{a}_3|\sqrt{ \frac{1}{3}-\frac{1}{2k^2}}$ and \newline
$\displaystyle D^*_{1\pm} = -\frac{2c\bar{a}_3}{\sqrt{3}}\pm\frac{2}{\sqrt{3}}\sqrt{1-4c^2-\frac{3}{2k^2}}$.
}
\label{Stability-table1}
\end{table}

We note that when $BI1^+_-$ is asymptotically stable, so is $BI1^+_+$.  This indicates that we can have two asymptotically stable centers within the $Q=1$ invariant set.  There exists ranges of parameter values for which there exists asymptotically stable periodic orbit(s) in a local neighbourhood of $BI1^+_\delta$ in the full phase space.  See \cite{VanDenHoogen:2018anx} for particular details on the existence of periodic orbits in the case $\bar{a}_3=0$.

\paragraph{Bianchi type I: Scalar Field Equilibrium Points (type 2) - $BI2^+_{\delta}$}

The two points $BI2^+_{\delta}$ can be expressed as
\begin{equation}
BI2^+_{\delta} =(\Phi_{eq},\Psi_{eq},y_{eq},1) =\left(\frac{\sqrt{2}}{k\tilde{a}_2}\left(\Psi_{eq}-\frac{\sqrt{2}}{\sqrt{3}}k\right),\Psi_{eq},\frac{\bar{a}_3}{\sqrt{2}ck\tilde{a}_2}\left(\Psi_{eq}-\frac{\sqrt{2}}{\sqrt{3}}k\right),1\right)
\end{equation}
where $\delta=\pm1$ and
\begin{equation}
\Psi_{eq}=\frac{k\left(\sqrt{2}(\bar{a}_3^{\,2}+4)+\delta2\tilde{a}_2\sqrt{\widetilde{K}}\right)}{\sqrt{3}(\bar{a}_3^{\,2}+4+2k^2\tilde{a}_2^{\,2})}.
\end{equation}
These equilibrium points represent anisotropic Bianchi type I models with a scalar field. These two points exist even when there is no coupling between the scalar field and the aether field and in that particular case represent the traditional isotropic power-law inflationary solutions when $k^2<\frac{1}{2}$.  If $\bar{a}_3=0$, then $BI2^+_{\delta}$ is the isotropic point $FR^+_{\delta}$ found in \cite{VanDenHoogen:2018anx}.

The points $BI2^+_\delta$ exist in the physical phase space if $\tilde{K}>0$ and $\Phi_{eq}\geq 0$.   The point $BI2^+_-$ exists in the physical phase space for parameter values
\begin{equation}
k^2\geq \frac{3}{2} \mbox{\qquad and\qquad} {\tilde a}_2<-\sqrt{\left({\bar a}_3^2+4\right)\left(\frac{1}{3}-\frac{1}{2k^2}\right)}.
\end{equation}
On the other hand, the point $BI2^+_+$ exists in the physical phase space if
\begin{eqnarray}	
&& k^2\leq \frac{3}{2} \mbox{\qquad  or }\\&& k^2\geq \frac{3}{2} \mbox{\qquad and\qquad }{\tilde a}_2<-\sqrt{\left({\bar a}_3^2+4\right)\left(\frac{1}{3}-\frac{1}{2k^2}\right)}.
\end{eqnarray}
See Figure \ref{Existence_plot} which summarizes graphically the existence of $BI1^+_\delta$ and $BI2^+_\delta$.

Since
\begin{equation}
\nabla f\rvert_{BI2^+_\delta}=[-2\Phi_{eq},-2\Psi_{eq},-2c^2y_{eq},0],
\end{equation}
the variable $\Phi$ can be eliminated locally at the equilibrium points. The eigenvalues for the resulting three dimensional system in the variables $\Psi,y,Q$ at the points $BI2^+_\delta$ in terms of $\Psi_{eq}$ are
\begin{equation}
\lambda_{1,2}=\sqrt{2}k\left(\Psi_{eq}-\frac{3}{\sqrt{6}k}\right),\quad\lambda_3= 2\sqrt{2}k\left(\Psi_{eq}-\frac{1}{\sqrt{6}k}\right)-\sqrt{\frac{2}{3}}\frac{\bar{a}_3}{ck\tilde{a}_2}\left(\Psi_{eq}-\frac{2k}{\sqrt{6}}\right)\label{eigensA+}
\end{equation}
The pair of repeated eigenvalues,
\begin{equation}
\lambda_{1,2}=\frac{2}{\sqrt{3}(2k^2\tilde{a}_2^{\,2}+\bar{a}_3^{\,2}+4)}\left( -\tilde{K}+\delta \mathrm{sgn}(\tilde{a}_2)\sqrt{\tilde{K}^2+\tilde{K}(2k^2\tilde{a}_2^{\,2}+\bar{a}_3^{\,2}+4)\left(k^2-\frac{3}{2}\right)}\right),
\end{equation}
correspond to the eigen-directions spanning the $Q=+1$ invariant set. It is straightforward to conclude that the two eigenvalues are positive when $BI2^+_-$ exists.  Similarly, the two eigenvalues are negative when $BI2^+_+$ exists. Therefore, for dynamics restricted to the $Q=1$ invariant set, $BI2^+_-$ is unstable and $BI2^+_+$ is stable whenever they exist.

The third eigenvalue indicates stability with respect to changes in the curvature, i.e, away from the $Q=1$ invariant set. The third eigenvalue can be expressed as a rather complicated expression involving all the parameters
\begin{equation}
\lambda_3=\frac{1}{3\sqrt{3}c^2(2k^2\tilde{a}_2^{\,2}+\bar{a}_3^{\,2}+4)}\left(\beta+\delta\mathrm{sgn}(\alpha)\sqrt{\beta^2+\Delta}\right)
\end{equation}
where
\begin{eqnarray}
\alpha&=&2\sqrt{3}ck^2\tilde{a}_2-\bar{a}_3\\
\beta&=&-12c^2k^2\left(\left(\tilde{a}_2-\frac{\sqrt{3}}{6c}\bar{a}_3\right)^2-\left(1-\frac{1}{2k^2}\right)\left(\bar{a}_3^{\,2}+4\right)-\frac{\bar{a}_3^{\,2}}{12c^2}\right)\\
\Delta&=&72c^4k^2\left(2k^2\tilde{a}_2^{\,2}+\bar{a}_3^{\,2}+4\right)\left(6k^2-1\right)\left(\left(\tilde{a}_2-\frac{2\sqrt{3}}{3c(6k^2-1)}\bar{a}_3\right)^2\right.\nonumber\\
      &&\qquad -\frac{(2k^2-1)^2}{2k^2(6k^2-1)}\left(\bar{a}_3^{\,2}+4\right)
        -\frac{(2k^2-1)^2}{2k^2c^2(6k^2-1)^2}\bar{a}_3^{\,2}\left.\vphantom{\left(\tilde{a}_2-\frac{2\sqrt{3}}{3c(6k^2-1)}\bar{a}_3\right)^2}\right)
\end{eqnarray}
Each of the three expressions $\alpha=0$, $\beta=0$ and $\Delta=0$ traces a two parameter $(\bar{a}_3,c)$ family of curves in the $(k,\tilde{a}_2)$ parameter space and consequently determines the sign of $\lambda_3$.

For the equilibrium point $BI2^+_-$, it can be shown that $\lambda_3>0$ whenever this point exists in the phase space.  Therefore, this point is a source whenever it exists.

For the equilibrium point $BI2^+_+$, it can be shown that $\lambda_3<0$ in the parameter regions described in Table \ref{Stability-table} and hence the point is asymptotically stable. In all other regions, $\lambda_3>0$ and the point becomes a saddle with a one dimensional unstable manifold. The same tabular information is represented graphically in Figures \ref{Inflationary-Stability1} and \ref{Inflationary-Stability2}. We note, if
\begin{equation}
\bar{a}_3<0,\qquad 0<c^2<\frac{\bar{a}_3^{\,2}}{(\bar{a}_3^{\,2}+4)}
\end{equation}
then there exists a disconnected region of parameter space in which $\lambda_3<0$. See the left-bottom graph in Figure \ref{Inflationary-Stability2}.

\begin{table}
$$\arraycolsep=1.4pt\def\arraystretch{2.2}
\begin{array}{|>{\centering\arraybackslash$} p{3.0cm} <{$}   >{\centering\arraybackslash$} p{3.0cm} <{$} >{\centering\arraybackslash$} p{4.0cm} <{$} >{\centering\arraybackslash$} p{3.0cm} <{$} |}
\hline
\bar{a}_3 & c^2 & k^2 & \tilde{a}_2\\
\hhline{|====|}
\multicolumn{4}{|l|}{{\bf BI2^+_+}  \qquad (\delta=+1)}\\
\bar{a}_3>0 & 0 < c^2 < 1 & 0< k^2 \leq \frac{1}{6} & -\infty < \tilde{a}_2 < \infty \\
            &             & \frac{1}{6} < k^2 < \infty & -\infty < \tilde{a}_2 < D^*_{2-} \\
\hline
\bar{a}_3=0 & 0 < c^2 < 1 & 0< k^2 \leq \frac{1}{6} & -\infty < \tilde{a}_2 < \infty \\
            &             & \frac{1}{6} < k^2 < \infty & -\infty < \tilde{a}_2 < D^*_{2-} \\
\hline
\bar{a}_3<0 & \frac{\bar{a}_3^{\,2}}{\bar{a}_3^{\,2}+4} < c^2 < 1 & 0< k^2 < (k^*_2)^2                 & -\infty < \tilde{a}_2 < \infty \\
            &             & (k^*_2)^2\leq k^2 < \frac{1}{6}    & -\infty < \tilde{a}_2 < D^*_{2-} \\
            &             & (k^*_2)^2\leq k^2 < \frac{1}{6}    & D^*_{2+} < \tilde{a}_2 < \infty \\
            &             & \frac{1}{6} \leq k^2< \infty      & -\infty < \tilde{a}_2 < D^*_{2-} \\
\hline
\bar{a}_3<0 & 0 < c^2 \leq \frac{\bar{a}_3^{\,2}}{\bar{a}_3^{\,2}+4} & 0 < k^2 <\frac{1}{6}            & -\infty < \tilde{a}_2 < D^*_{2-} \\
            &             & 0 < k^2 <\frac{1}{6}            & D^*_{2+} < \tilde{a}_2 < \infty  \\
            &             & \frac{1}{6} \leq k^2 <\infty   & -\infty < \tilde{a}_2 < D^*_{2-} \\
\hline
\end{array}$$
\caption{Regions of the $(k,\tilde{a}_2)$ parameter space where the equilibrium point $BI2^+_-$ is stable. \newline
$\displaystyle (k^*_2)^2 = \frac{1}{6c^2}\left(c^2-\frac{\bar{a}_3^{\,2}}{\bar{a}_3^{\,2}+4}\right)$ and
$\displaystyle D^*_{2\pm} = \frac{4\sqrt{3}k\bar{a}_3\pm3\sqrt{2}(2k^2-1)\sqrt{\bar{a}_3^{\,2}+(6k^2-1)(\bar{a}_3^{\,2}+4)}}{6kc(6k^2-1)}$
.}
\label{Stability-table}
\end{table}

\begin{figure}[ht]
\begin{center}\arraycolsep=1.4pt\def\arraystretch{2.2}
\begin{tabular}{|c c|}
\hline
\multicolumn{2}{|c|}{$\displaystyle  \bar{a}_3>0,\quad 0<c^2\leq 1$ }\\
\includegraphics[width=0.4\textwidth]{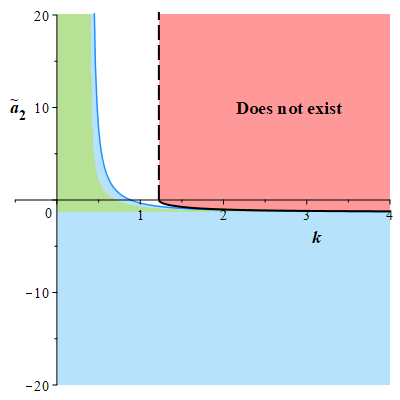} &
\includegraphics[width=0.4\textwidth]{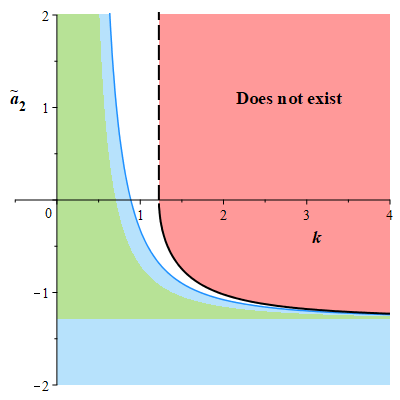}\\
\hline
\multicolumn{2}{|c|}{$\displaystyle  \bar{a}_3=0,\quad 0<c^2\leq 1$} \\
\includegraphics[width=0.4\textwidth]{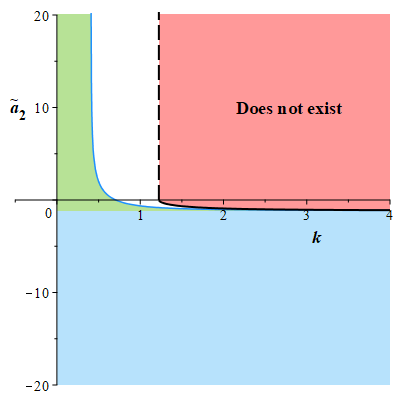}&
\includegraphics[width=0.4\textwidth]{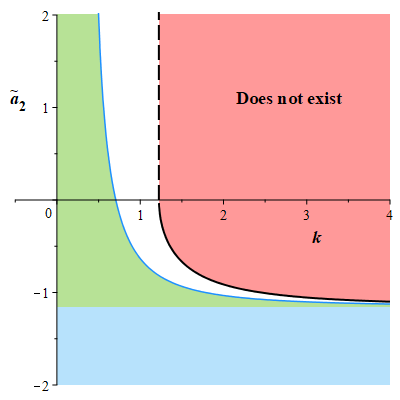}\\
\hline
\end{tabular}
\end{center}
\caption{A wide and a vertically scaled view of stability for the equilibrium point $BI2^+_+$ when $\bar{a}_3\geq0$. The blue and green regions indicate values in the $(k,\tilde{a}_2)$ parameter space in which the equilibrium point $BI2^+_+$  is an attractor.  Yellow and green areas indicate parameter values where the models are expanding with $q<0$, that is inflationary.  Green areas yield parameter values in which there is a stable inflationary attractor. The values of $\bar{a}_3$ and $c$ were chosen to highlight the shapes of the different regions. }\label{Inflationary-Stability1}
\end{figure}

\begin{figure}[ht]
\begin{center}\arraycolsep=1.4pt\def\arraystretch{2.2}
\begin{tabular}{|c c|}
\hline
\multicolumn{2}{|c|}{$\displaystyle  \bar{a}_3<0,\quad \frac{\bar{a}_3^{\,2}}{\bar{a}_3^{\,2}+4}<c^2\leq 1$} \\
\includegraphics[width=0.4\textwidth]{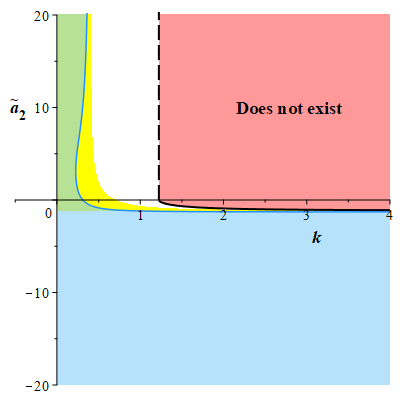}&
\includegraphics[width=0.4\textwidth]{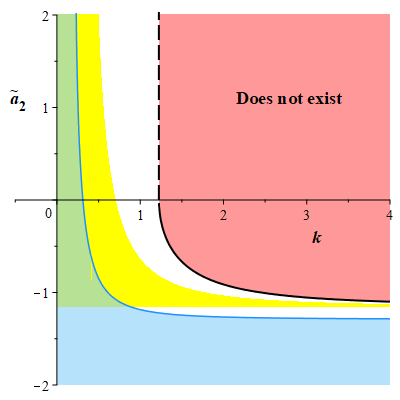}\\
\hline
\multicolumn{2}{|c|}{$\displaystyle  \bar{a}_{3}<0,\quad 0<c^2\leq \frac{\bar{a}_{3}^2}{\bar{a}_{3}^{\,2}+4}$} \\[5pt]
\includegraphics[width=0.4\textwidth]{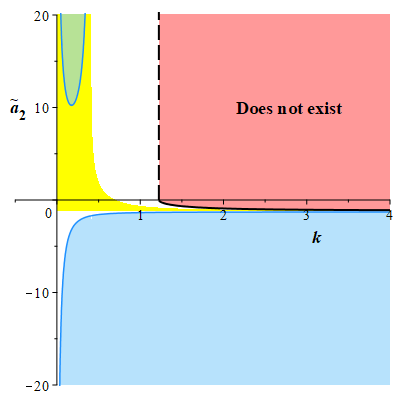}&
\includegraphics[width=0.4\textwidth]{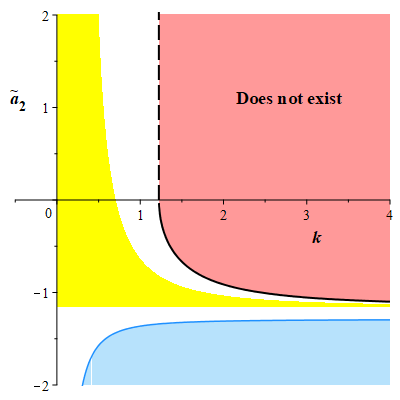}\\
\hline
\end{tabular}
\end{center}
\caption{A wide and a vertically scaled view of stability for the equilibrium point $BI2^+_+$ when $\bar{a}_3<0$. The blue and green regions indicate values in the $(k,\tilde{a}_2)$ parameter space in which the equilibrium point $BI2^+_+$  is an attractor.  Yellow and green areas indicate parameter values where the models are expanding with $q<0$, that is inflationary.  Green areas yield parameter values in which there is a stable inflationary attractor. The values of $\bar{a}_3$ and $c$ were chosen to highlight the shapes of the different regions. }\label{Inflationary-Stability2}
\end{figure}

\subsection{Power-Law Inflation}

The addition of a nontrivial coupling between the scalar field and the aether field changes the parameter requirements for power-law inflation to occur.  The only equilibrium point present in the constructed model for which power-law inflation could occur is the point $BI2^+_+$.  At this point
\begin{equation}
\tilde{q}\rvert_{BI2_+^+}=\sqrt{6}k\Psi_{eq}-1\label{inflationA+},
\end{equation}
which must be negative during inflation.  Further, at this equilibrium point we can simplify equation \eqref{def_q} and integrate
\begin{equation}
\frac{\dot \theta}{\theta^2}=-\frac{\sqrt{6}k}{3}\Psi_{eq} \Rightarrow \theta=\frac{3}{\sqrt{6}k\Psi_{eq}}(t-t_0)^{-1},
\end{equation}
where we see that $\Psi_{eq}>0$ else the model is contracting.  We observe that $\Psi_{eq}>0$ corresponds to the parameter constraint $\tilde{a}_2>-\frac{1}{\sqrt{3}}\sqrt{\bar{a}_3^{\,2}+4}$. Therefore this equilibrium point represents an expanding power-law inflationary solution if
\begin{equation}
0 <\sqrt{6}k\Psi_{eq} < 1.
\end{equation}
These constraints correspond to the parameter region bounded between
\begin{eqnarray}
k^2\leq\frac{1}{6},\qquad && -\sqrt{\frac{\bar{a}_3^{\,2}+4}{3}}<\tilde{a}_2<\infty  \\
k^2>\frac{1}{6},\qquad    && -\sqrt{\frac{\bar{a}_3^{\,2}+4}{3}}<\tilde{a}_2< -\frac{(2k^2-1)}{2k^2}\sqrt{\frac{(\bar{a}_3^{\,2}+4)2k^2}{(6k^2-1)}}. \nonumber
\end{eqnarray}
If $\bar{a}_3=0$ then we obtain the lower bound for power-law inflation to occur to be $\tilde{a}_2>-2/\sqrt{3}$.  This corrects an incorrect observation in \cite{VanDenHoogen:2018anx} which concluded that the lower bound for inflation is $\tilde{a}_2=-\infty$,  indeed, inflation cannot occur if $\tilde{a}_2<-2/\sqrt{3}$. See green shaded regions in Figures \ref{Inflationary-Stability1} and \ref{Inflationary-Stability2}.

\section{Numerical Analysis and Qualitative Summary}\label{Numerical}

We confirm the qualitative analysis of the previous section with a few numerical calculations.  The values of the parameters are selected to highlight the different asymptotic behaviours that are possible.

\subsection{Bianchi I (\texorpdfstring{$Q=1$}{Q=1}) Invariant Set}

\begin{figure}[ht]
\begin{center}
	$\begin{array}{rl}
	\includegraphics[width=0.4\textwidth]{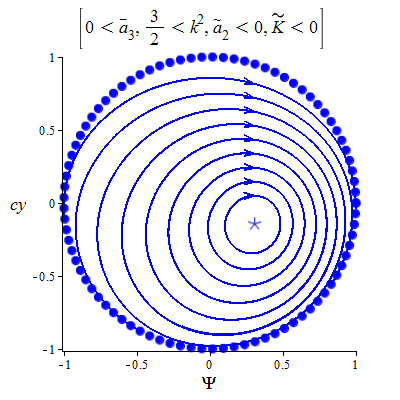}  &
	\includegraphics[width=0.4\textwidth]{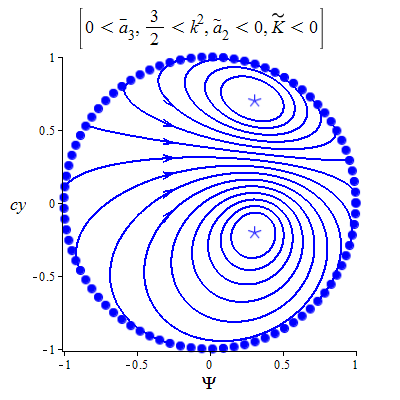} \\
	\includegraphics[width=0.4\textwidth]{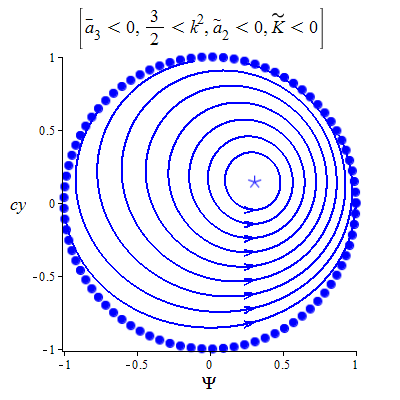}  &
	\includegraphics[width=0.4\textwidth]{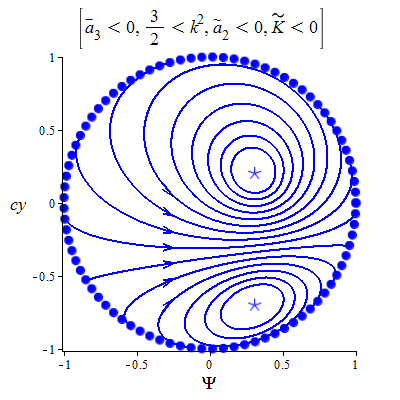}
	\end{array}$
\end{center}
	\caption{The figures depict phase portraits of typical behaviours when $\bar{a}_3>0$ and $\bar{a}_3<0$ for the $BI1^+_\delta$ equilibrium points in the $Q=1$ invariant set. The solid circles represent the $C^+$ circle of equilibrium points, while an asterisk represents the points $BI1^+_\delta$. Note how $\bar{a}_3$ tilts the figures.  }\label{BI stability1}
\end{figure}

\begin{figure}[ht]
\begin{center}
	$\begin{array}{rl}
	\includegraphics[width=0.4\textwidth]{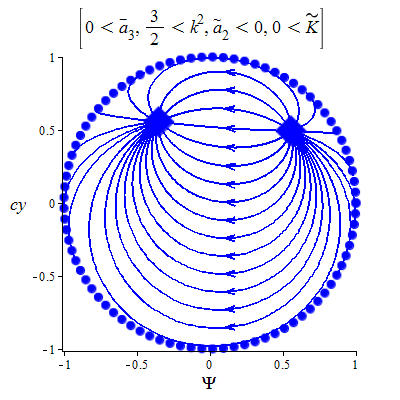}  &
	\includegraphics[width=0.4\textwidth]{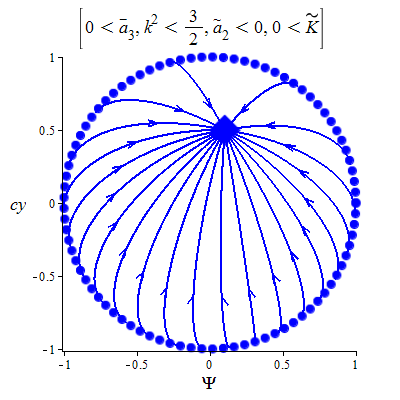} \\
	\includegraphics[width=0.4\textwidth]{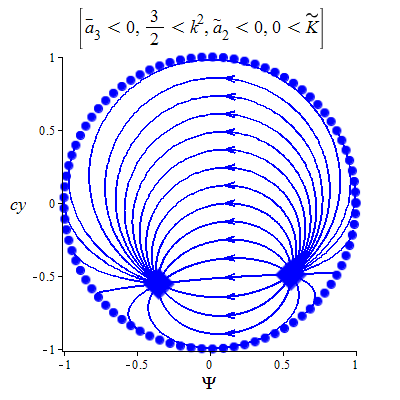}  &
	\includegraphics[width=0.4\textwidth]{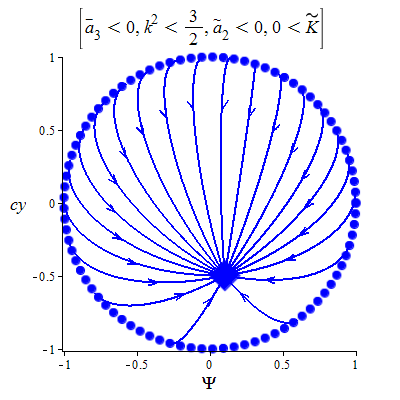}
	\end{array}$
\end{center}
	\caption{The figures depict phase portraits of typical behaviours when $\bar{a}_3>0$ and $\bar{a}_3<0$ for the $BI2^+_\delta$ equilibrium points in the $Q=1$ invariant set. The solid circles represent the $C^+$ circle of equilibrium points, while solid diamonds represent the points $BI2^+_\delta$. Note how $\bar{a}_3$ tilts the figures.  }\label{BI stability2}
\end{figure}

In the $Q=1$ invariant set the dynamics are dominated by the different possible points: $C^+$, $BI1^+_\delta$ and $BI2^+_\delta$. While $BI1^+_\delta$ and $BI2^+_\delta$ cannot occur simultaneously,  it is possible for both $BI1^+_-$ and $BI1^+_+$ (or $BI2^+_-$ and $BI2^+_+$) to be simultaneously present.  See Figures \ref{BI stability1} and \ref{BI stability2}.  Note for certain values of the parameters, neither  $BI1^+_\delta$ nor $BI2^+_\delta$ exist, in which case the phase portrait evolves from a subset of points on $C^+$ to different subset of points on $C^+$.  A figure has not been included.

In the Bianchi I invariant set, if $k^2<\frac{3}{2}$ then the unique future asymptotic state is represented by the equilibrium point $BI2^+_+$.  If $k^2>\frac{3}{2}$ and $\tilde{a_2}<\tilde{K}_-$ then the future asymptotic state is either $BI2^+_+$ or a subset of points in $C^+$. In addition to the asymptotically stable points, families of periodic orbits are possible. Define
\begin{equation}
\tilde{K}_\pm=\pm\sqrt{(\bar{a}_3^{\,2}+4)\left(\frac{1}{3}-\frac{1}{2k^2}\right)},\qquad \tilde{\tilde{K}}_\pm=\pm|\bar{a}_3|\sqrt{ \left(\frac{1}{3}-\frac{1}{2k^2}\right)}.
\end{equation}
If $k^2>\frac{3}{2}$ and $\tilde{K}_-<\tilde{a}_2<\tilde{\tilde{K}}_-$ then there are two different families of periodic orbits within the Bianchi I invariant set. If $k^2>\frac{3}{2}$ and $\tilde{\tilde{K}}_-<\tilde{a}_2<\tilde{\tilde{K}}_+$ then there is a single family of periodic orbits (See Figures \ref{BI stability1} and \ref{BI stability2}).

\subsection{Kantowski-Sachs \texorpdfstring{($Q^2<1$)}{Qless1} Invariant Set}

\begin{figure}[ht]
\begin{center}
	$\begin{array}{cc}
     c=\frac{ 2}{32},\bar{a}_3=\sqrt{4},k=16,\tilde{a}_2=-\frac{9}{10}\\
	\includegraphics[width=0.4\textwidth]{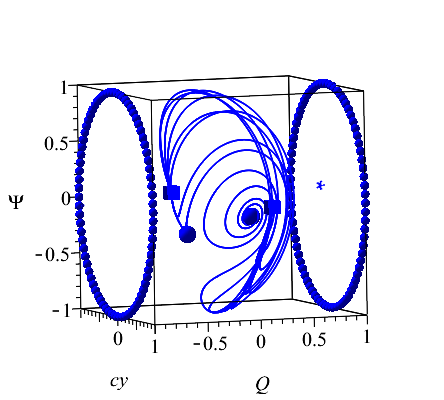}  &
    \includegraphics[width=0.4\textwidth]{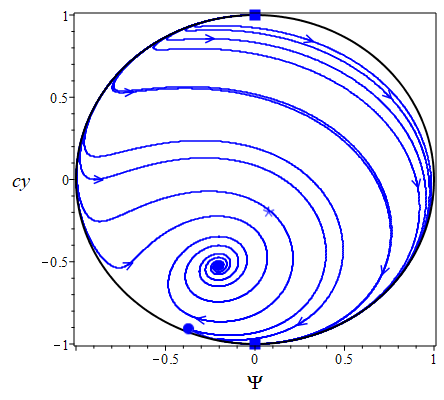}
	\end{array}$
\end{center}
	\caption{The figures depicts a three-dimensional phase portrait and its two-dimensional projection.   The equilibrium points $KS2_-$ and $KS2_+$ are indicated by solid circles.  The asterisk represent the $BI1^+_\delta$ and $BI1^-_\delta$ equilibrium points, while the circle of equilibrium points in the right-hand figure represent $C^-$ and $C^+$. The figures illustrates how the equilibrium point $KS2_+$ is an asymptotically stable equilibrium point. }\label{KS2 stability_phase}
\end{figure}

In the three dimensional phase space $Q^2<1$, the number of equilibrium points depends on the value of parameters.  When $c<\frac{1}{2}$ then there are at least two equilibrium points and as many as four different equilibrium points are possible. When $c>\frac{1}{2}$ then there can be two equilibrium points and it is possible that there are none.  Different three dimensional phase portraits can be constructed, but their value is limited.  We include one figure that illustrates a situation in which there are two attractors in the three dimensional phase space.  The asymptotic stable attractors are the equilibrium points $KS2_+$ and $KS1_-$, the source is $KS1_+$, while the point $KS2_-$ is a saddle (see Figure \ref{KS2 stability_phase}). There are other attractors in the $Q=1$ invariant set that are not shown.

In the Kantowski-Sachs invariant set, the future asymptotic attractor can be one of the following $C^+$, $BI1^+_+$, $BI1^+_-$, $BI2^+_+$, $KS1_-$, and $KS2_+$. Indeed, it is also possible for one or two families of asymptotically stable periodic orbits. The point $BI2^+_+$ is the point of most interest as this equilibrium point represents an expanding inflationary solution for appropriate values of the parameters.  For different parameter values, this equilibrium point is also asymptotically stable.  We are of course interested in when this point represents an asymptotically stable expanding inflationary solution. The green areas in Figures \ref{Inflationary-Stability1} and \ref{Inflationary-Stability2} represent possible parameter values where an asymptotically stable expanding inflationary solution exists.

\section{Concluding Remarks}

Investigating Einstein-Aether theories allows us to consider non-trivial couplings between the aether field and the scalar field which are not possible in General Relativity. Such couplings can result in altered stabilities of the late-time power-law inflationary solution, however, in \cite{VanDenHoogen:2018anx}, it was found that the presence of a non-trivial coupling between the expansion of the scalar and aether fields preserved the stability of the isotropic power-law inflationary solution. This stability was shown to persist even for values of the scalar field potential parameter $k^2>\frac{1}{2}$ as long as an appropriate value for the expansion couple parameter $a_2$ is chosen.

In this paper, we coupled the scalar field directly to the aether field via both its expansion and shear, and found changes in the future asymptotic dynamics resulting from the coupling of the scalar field to the shear of the aether field. Chiefly, we observed that the previously isotropic expansionary power-law inflationary attractor became anisotropic and is represented here by the equilibrium point $BI2^{+}_{+}$. In accordance with \cite{VanDenHoogen:2018anx}, the solution corresponding to this equilibrium point is only isotropic in the absence of shear coupling ($a_3=0$). Further, the stability of the expansionary power-law inflationary solution changes depending on the value of $a_3$, with the solution (when it exists) being stable for values of $a_3>0$ (see green regions in Figures \ref{Inflationary-Stability1}), and unstable for values of $a_3<0$ (see yellow regions in Figures \ref{Inflationary-Stability2}), for some values of the parameters ($k,a_2$).

We also note that the non-trivial coupling between the scalar field and aether field brings into existence, for a particular range of parameter values, a family of asymptotically stable periodic orbits. The solutions represented by these orbits are spatially homogeneous but anisotropic, and have a scalar field with an exponential potential. It is also worth noting that these families of solutions do not exist in the absence of a coupling between the scalar and aether fields.

Evidently, violating the boost part of the Lorentz invariance within both the geometry and matter sectors of a gravitational theory results in dynamics different from what one would expect. In this investigation, we sought not only to further illuminate that assertion, but to specify which qualitative changes become possible with the particular Lorentz non-invariances we have prescribed. Changes we found as a result of these violations include: the existence of future asymptotic states that are periodic in nature, expansionary power-law inflationary solutions that are not future asymptotically stable, and expansionary power-law inflationary solutions that are not necessarily isotropic.

%% ----------------------------------------------------------------
%% --     End Matter                -------------------------------
%% ----------------------------------------------------------------

\acknowledgments SM acknowledges the support of NSERC through an Undergraduate Student Research Award (USRA). RvdH and DW were supported by the St. Francis Xavier University Council on Research.

\bibliographystyle{JHEP}
\bibliography{Reference_Aether}

\clearpage

\appendix
\section{The  \texorpdfstring{$KS2_\delta$}{KS2delta}  Equilibrium Points}\label{appendixa}
The Kantowski-Sachs points $KS2_\delta$ have the general form
$$[\Phi,\Psi,y,Q] = [\Phi^*,\Psi^*,y^*,Q^*]$$
where $\Psi^*$ and $Q^*$ are the following linear combinations of $\Phi^*$ and $y^*$.
\begin{eqnarray}
\Psi^*&=&\frac{-\sqrt{6}k(\sqrt{3}\tilde{a}_2+2\bar{a}_3 c)}{6(2k^2-1)}\Phi^*+\frac{\sqrt{6}k(2c^2+1)}{3(2k^2-1)}y^*\\
Q^*&=&\frac{-k^2(\sqrt{3}\tilde{a}_2+2\bar{a}_3 c)}{(2k^2-1)}\Phi^*+\frac{(4c^2k^2+1)}{(2k^2-1)}y^*
\end{eqnarray}
The above system becomes singular when $k^2=\frac{1}{2}$.  The value of $y^*$ and $\Phi^*$ can be determined via the following coupled expressions where $\delta=\pm 1$,
\begin{eqnarray}
\Phi^* &=& \sqrt{\frac{2A_1C_2}{B_1B_2-2A_1A_2-\delta B_2 \sqrt{D_1}}} \label{phi_eq}\\
y^* &=& \frac{-B_1+\delta\sqrt{D_1}}{2A_1} \Phi^*
\end{eqnarray}
where the values of $A_i,B_i,C_i,D_i$ are functions of the various parameters $(c,k,\tilde{a}_2,\bar{a}_3)$ in the model
\begin{eqnarray*}
	A_1 &=& -12 (2{c}^{2}+1) \left( 8c^2k^2-2k^2+3 \right)
	\\
	B_1 &=& 6c \left( 32{c}^{2}{k}^{2}+12{k}^{4}-8{k}^{2}+9 \right) {\bar{a}_3}+24\sqrt {3} \left( 6{c}^{2}{k}^{2}+{c}^{2}+2 \right) \tilde{a}_2{k
	}^{2}
	\\
	C_1 &=& -48{c}^{2}{k}^{2}{\bar{a}_3}^{2}-12\sqrt {3}c \left( 6{k}^{2}+1
	\right) {k}^{2}\bar{a}_3\tilde{a}_2-18{k}^{2}\left(6k^2-1\right){\tilde{a}_2}^{2}+36\left(2k^2-1\right)^2
	\\
	D_1 &=& (B_1)^2-4A_1C_1\\
	A_2 &=& -4{c}^{2}{k}^{2} \left( 4{c}^{2}{k}^{2}+1 \right) {\bar{a}_3}^{2}-2
	\sqrt {3} \left( 8{c}^{4}{k}^{2}+12{c}^{2}{k}^{4}+4{c}^{2}{k}^
	{2}+3{c}^{2}+2{k}^{2}+2 \right) c{k}^{2}\bar{a}_3\tilde{a}_2
	%% line break
	\\ &&\qquad
	-3k^2\left( 8k^2c^4+12c^2k^4+3c^2+2k^2+1\right){\tilde{a}_2}^{2}
	\\ && \qquad
	+6\left(2k^2-1\right)\left(8k^2c^4+4c^2k^4+3c^2+1\right)
	\\
	%% line break
	%%\right.\\ &&\qquad \left.
	%%
	B_2 &=& c \left( 80{c}^{4}{k}^{4}+24{c}^{2}{k}^{6}-24{c}^{4}{k}^{2}-12
	{c}^{2}{k}^{4}+34{c}^{2}{k}^{2}+4{k}^{4}-9{c}^{2}+2{k}^{2}
	\right) \bar{a}_3
	\\ &&\qquad +4\sqrt {3} \left( 8{k}^{2}{c}^{6}+12{c}^{4}{k}
	^{4}+4{c}^{4}{k}^{2}+3{c}^{4}+6{c}^{2}{k}^{2}+2{c}^{2}+1
	\right){k}^{2}\tilde{a}_2
	\\
	C_2 &=& -2(2k^2-1) (2{c}^{2}+1) \left(8c^2k^2-2k^2+3\right)
\end{eqnarray*}
This pair of points only exist provided $D_1\geq0$ and equation \eqref{phi_eq} yields a real root for $\Phi^*$.  The equilibrium point is within the physical phase-space provided $(Q^*)^2\leq 1$.

\clearpage
\end{document}